\documentclass[twocolumn,aps,showpacs,preprintnumbers,byrevt 
ex,superscriptaddress]{revtex4} 
\usepackage{graphicx} 
\usepackage{dcolumn} 
\usepackage{bm}

\begin{document}

\title{Quantum correlations in bound-soliton pairs and trains in 
fiber lasers} 
\author{Ray-Kuang Lee} 
\affiliation{Department of Photonics and Institute of 
Electro-Optical Engineering, National Chiao-Tung University, Hsinchu 
300, Taiwan} 
\author{Yinchieh Lai} 
\email{yclai@mail.nctu.edu.tw} 
\affiliation{Department of Photonics and Institute of Electro-Optical Engineering, National 
Chiao-Tung University, Hsinchu 300, Taiwan} 
\author{Boris A. Malomed} 
\email{malomed@eng.tau.ac.il} 
\affiliation{Department of 
Interdisciplinary Studies, Faculty of Engineering, Tel Aviv 
University, Tel Aviv 69978, Israel} 
\date{\today}

\begin{abstract} 
Quantum correlations in pairs and arrays (trains) 
of bound solitons modeled by the complex Ginzburg-Landau equation (CGLE) are 
calculated numerically, on the basis of linearized equations for 
quantum fluctuations. We find strong correlations between the bound 
solitons, even though the system is dissipative. Some degree of the 
correlation between the photon-number fluctuations of stable 
bound soliton pairs and trains is attained and saturates after 
passing a certain distance. The saturation of the photon- number 
correlations is explained by the action of non- conservative terms 
in the CGLE. Photon-number-correlated bound soliton trains offer 
novel possibilities to produce multipartite entangled sources for 
quantum communication and computation. 
\end{abstract}

\maketitle

Solitons in optical fibers are well known for their remarkable 
dynamical properties, at both classical and quantum levels. As 
concerns the latter, the solitons are macroscopic optical fields 
that exhibit quadrature-field squeezing of quantum fluctuations 
around the classical core \cite{Drummond87, Lai89, Lai90}, as well 
as amplitude squeezing \cite{Friberg, RK-fbg}, and both intra-pulse 
and inter-pulse correlations \cite{Schmidt00}. Due to the Kerr 
nonlinearity of the fiber, the quantum fluctuations about the 
temporal solitons get squeezed during the propagation, i.e., the 
variance of the perturbed quadrature field operator around the 
soliton is \emph{smaller} than in the vacuum state. 
The nonlinear Schr{\"{o}}dinger equation (NLSE) is a commonly 
adopted model for the description of classical and quantum dynamics 
in optical fibers. Experimental investigations of quantum properties 
of temporal solitons have shown remarkable agreement with 
predictions of the NLSE, provided that loss and higher-order effects 
are negligible \cite{Friberg, Rosenbluh, Bergman91, Yu01, Krylov99, 
Spalter}.

In the NLSE, adjacent temporal solitons attract or repel each other, 
depending on the phase shift between them \cite{Agrawal95}. In the 
cases when the potential interaction force between the solitons can 
be balanced by additional effects, such as those induced by small 
loss and gain terms in a perturbed cubic \cite{Malomed91,Jena} or 
quintic NLSE \cite{Akhmediev96, Akhmediev97, Soto97, 
Seva,Akhmediev98} [which actually turns the NLSE equation into a 
complex Ginzburg-Landau equation (CGLE)], or the polarization 
structure of the optical field, described by the coupled NLSEs 
\cite{Haelterman93, Kaup93, Malomed98}, bound states of solitons 
have been predicted. Recently, formation of stable double-, triple-, 
and multi-soliton bound states (\textit{trains}, in the latter case) 
has been observed experimentally in various passively mode-locked 
fiber-ring laser systems \cite{Tang01, Seong02,Grelu03}, which 
offers potential applications to optical telecommunications. 
Formation of ``soliton crystals" in nonlinear fiber rings has been 
predicted too \cite{Fedor}.

Multiple-pulse generation in the passively mode-lock fiber lasers is 
quite accurately described by the quintic CGLE (which is written 
here in a normalized form), \begin{eqnarray} 
iU_{z}+\frac{D}{2}U_{tt}+|U|^{2}U &=&i\delta U+i\epsilon 
|U|^{2}U+i\beta U_{tt}  \nonumber \\ &+&i\mu |U|^{4}U-v|U|^{4}U, 
\label{eq_CGLE}
\end{eqnarray}
where $U$ is the local amplitude of 
the electromagnetic wave, $z$ is the propagation distance, $t$ is 
the retarded time, and $D$ corresponds to anomalous dispersion 
($+1$) or normal dispersion ($-1$). Besides the group-velocity 
dispersion (GVD)\ and Kerr effect, which are accounted for by 
conservative terms on the left-hand-side of Eq. (\ref{eq_CGLE}), the 
equation also includes the quintic correction to the Kerr effect, 
through the coefficient $\nu $, and non-conservative terms (the 
coefficients $\delta$, $\epsilon$, $\mu$, and $\beta$ account for 
the linear, cubic, and quintic loss or gain, and spectral filtering, 
respectively).

In quantum-squeezing experiments, additional noises due to the 
processes other than the GVD\ and Kerr effect, such as the acoustic-
wave Brillouin scattering, are unwanted and suppressed, using stable 
fiber lasers \cite{Yu01}. Accordingly, the non-conservative terms in 
the CGLE may be superficially considered as detrimental to the 
observation of quantum fluctuations of fiber solitons. However, 
an accurate analysis of the quantum fluctuations in the CGLE- based 
model is necessary, and was missing thus far, to the best of our 
knowledge.

The objective of the present work is to calculate quantum 
fluctuations around bound states of solitons in the CGLE by dint of 
a numerically implemented \textit{back-propagation method} 
\cite{Lai95}. We find strong quantum-perturbation correlations 
between the bound solitons, despite the fact the dissipative nature 
of the model will indeed prevent observation of the squeezing of the 
quantum fluctuations around the bound solitons in the fiber-ring 
lasers described by the CGLE. Multimode quantum-correlation spectra 
of the bound-soliton pairs show patterns significantly different 
from those for two-soliton configurations in the conservative NLSE. 
We also find that a similarity in the photon-number correlations 
between the stable bound-soliton pairs and multi-soliton trains.

Following the known approach to the investigation of bound- soliton 
states \cite{Malomed91,Akhmediev96,Seva}, the corresponding solution 
to Eq. (\ref{eq_CGLE}) is sought for in the form \[ 
U(z,t)=\sum_{j=1}^{N}U_{0}(z,t+\rho _{j})e^{i\theta _{j}}, \] where 
$U_{0}$ is a single-soliton solution, and $\rho _{j}$ and $\theta 
_{j}$ are the coordinates separation and phases of the individual 
solitons. Through the balance between the gain and loss, in- phase 
and out-of-phase bound-soliton pairs may exist in the anomalous-GVD 
regime, which is described by Eq. (\ref{eq_CGLE}). In the case of 
the normal dispersion, which corresponds to the opposite sign in 
front of $U_{tt}$ on the left-hand side of Eq. (\ref{eq_CGLE}), 
strongly chirped solitary pulses (which differs them from the 
classical solitons) and their bound states are possible too.

To evaluate the quantum fluctuations around the bound solitons, we 
replace the classical function $U(z,t)$ in Eq. (\ref{eq_CGLE}) by 
the quantum-field operator variable, $\hat{U}(z,t)$, which satisfies 
the equal-coordinate Bosonic commutation relations. Next we 
linearize the equation around the classical solution, i.e. 
$\hat{U}(z,t) = U_{0} + \hat{u}(z,t)$, for a state containing a very 
large number of photons. Then, the above-mentioned back-propagation 
method is used to calculate the perturbed quantum fluctuations 
around the bound-soliton states in the full CGLE model. The 
linearized equation for the perturbed field operator $\hat{u}(z,t)$ 
is, 
\begin{eqnarray} 
\frac{d}{d z} \hat{u}(z,t) = {\cal P}_1 (z,t) 
\hat{u}(z,t) + {\cal P}_2 (z,t) \hat{u}^\dag(z,t) + \hat{n}(z,t), 
\label{lineq} 
\end{eqnarray} 
where ${\cal P}_1$ and ${\cal P}_2$ are 
two special operators defined as follows, 
\begin{eqnarray*} {\cal 
P}_1(z,t) &=& i\frac{D}{2}\frac{\partial ^2}{\partial t^2} + 2 i 
|U_0|^2 + \delta +2 \epsilon |U_0|^2 + \beta 
\frac{\partial^2}{\partial t^2}\\ &+& 3 \mu |U_0|^4 +3 i \nu 
|U_0|^4,\\ {\cal P}_2(z,t) &=& i U_0^2 + \epsilon U_0^2  + 2 \mu 
U_0^3 U_0^\ast +2 i \nu U_0^3 U_0^\ast. 
\end{eqnarray*} 
To satisfy 
the Bosonic communication relations for the perturbed quantum fields 
$\hat{u}(z,t)$ and $\hat{u}^\dag (z,t)$, we also introduce a zero-mean additional noise operator $\hat{n}(z,t)$ in Eq. (\ref{lineq}), 
with satisfy following commutation relations \cite{Lai95}, 
\begin{eqnarray*} 
[\hat{n}(z,t_1)&,& \hat{n}^\dag(z',t_2)]= \\ && 
\{-{\cal P}_1 (z, t_1) - {\cal P}_1^\ast (z', t_2)\}\delta(z-
z')\delta(t_1-t_2),\\ \nonumber [\hat{n}(z,t_1)&,& \hat{n}(z',t_2)]= 
[\hat{n}^\dag(z,t_1), \hat{n}^\dag(z',t_2)]= 0. 
\end{eqnarray*}

To actually determine the correlation functions of $\hat{n}(z,t)$ and 
$\hat{n}^\dag(z,t)$, one has to consider its physical origins. In general,
$\hat{n}(z,t) = \sum_i \hat{n}_i(z,t)$, with $\hat{n}_i(z,t)$ being the noise 
operator contributed by the i-th non-conservative term in the equation. 
The commutation relations of $\hat{n}_i(z,t)$ and $\hat{n}_i^\dag(z,t)$ are of 
the same form as in the above equations for $\hat{n}(z,t)$ and $\hat{n}^\dag(z,t)$, 
except that the differential operators $P_1(z,t)$ and $P_2(z,t)$ contain only the 
corresponding non-conservative term.   
For simplicity, in our calculation we will introduce the following assumptions: 
$\langle \hat{n}_i^\dag(z,t_1) \hat{n}_i(z',t_2) \rangle = 0$ for loss terms
and $\langle \hat{n}_i(z,t_1) \hat{n}_i^\dag(z',t_2) \rangle = 0$ for gain terms. 
With these additional assumptions, the correlation functions for each $\hat{n}_i(z,t)$ 
as well as for the total noise $\hat{n}(z,t)$ can be calculated from their commutation relations.  
Physically, these assumptions are equivalent to assume the reservoirs corresponding to the 
loss terms are in the ground state and the population inversions corresponding to the gain 
terms are in full inversion. The magnitude of the noise level calculated with these assumptions 
represents the minimum quantum noise that will be introduced with the presence of the considered 
non-conservative terms. For real systems, the actual introduced noises will be always larger and thus    
our calculation results here only represent the lower bound limit required by the fundamental quantum 
mechanics principles.

\begin{figure}[tbp] 
\begin{center} 
\includegraphics[width=3.0in]{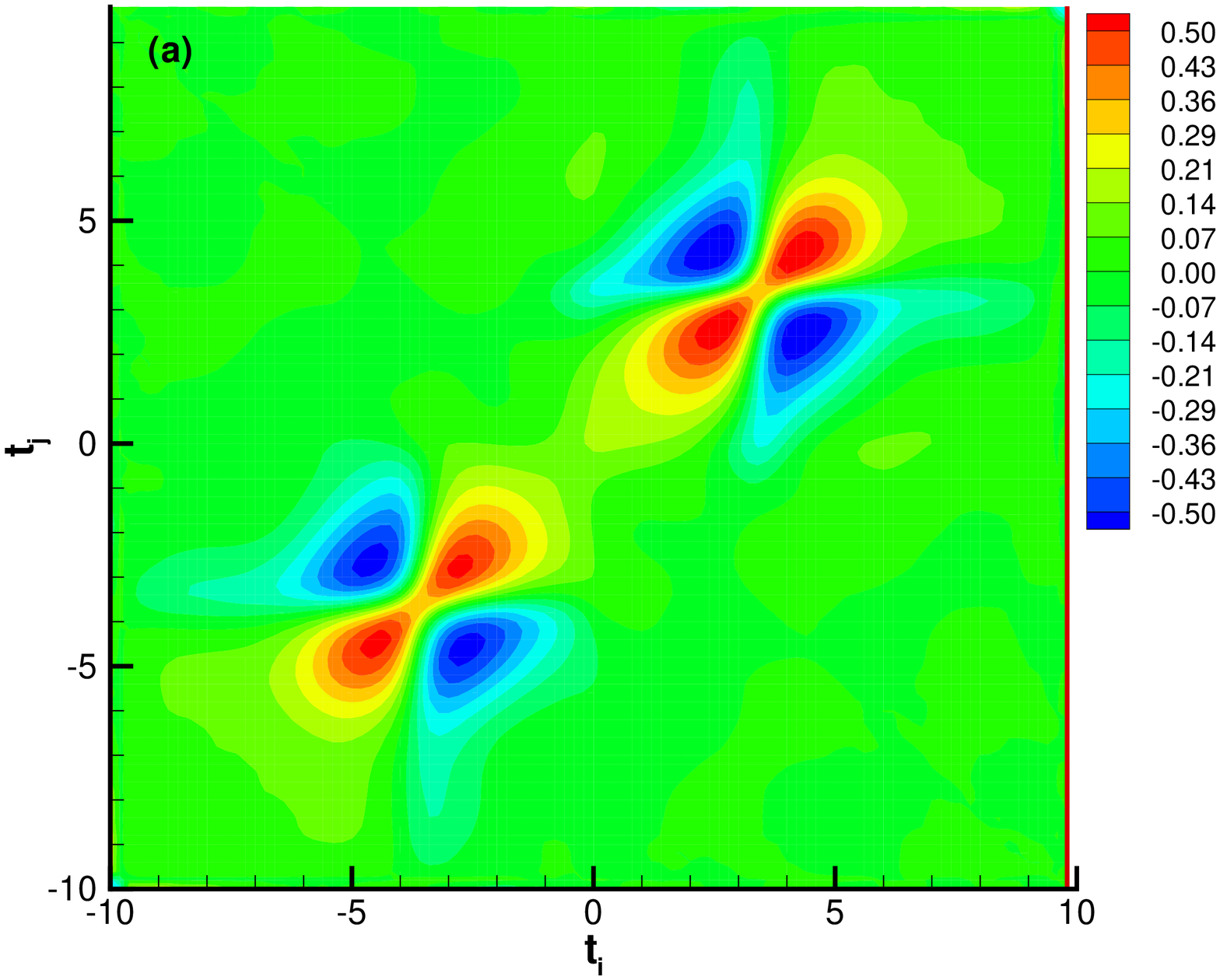} 
\includegraphics[width=3.0in]{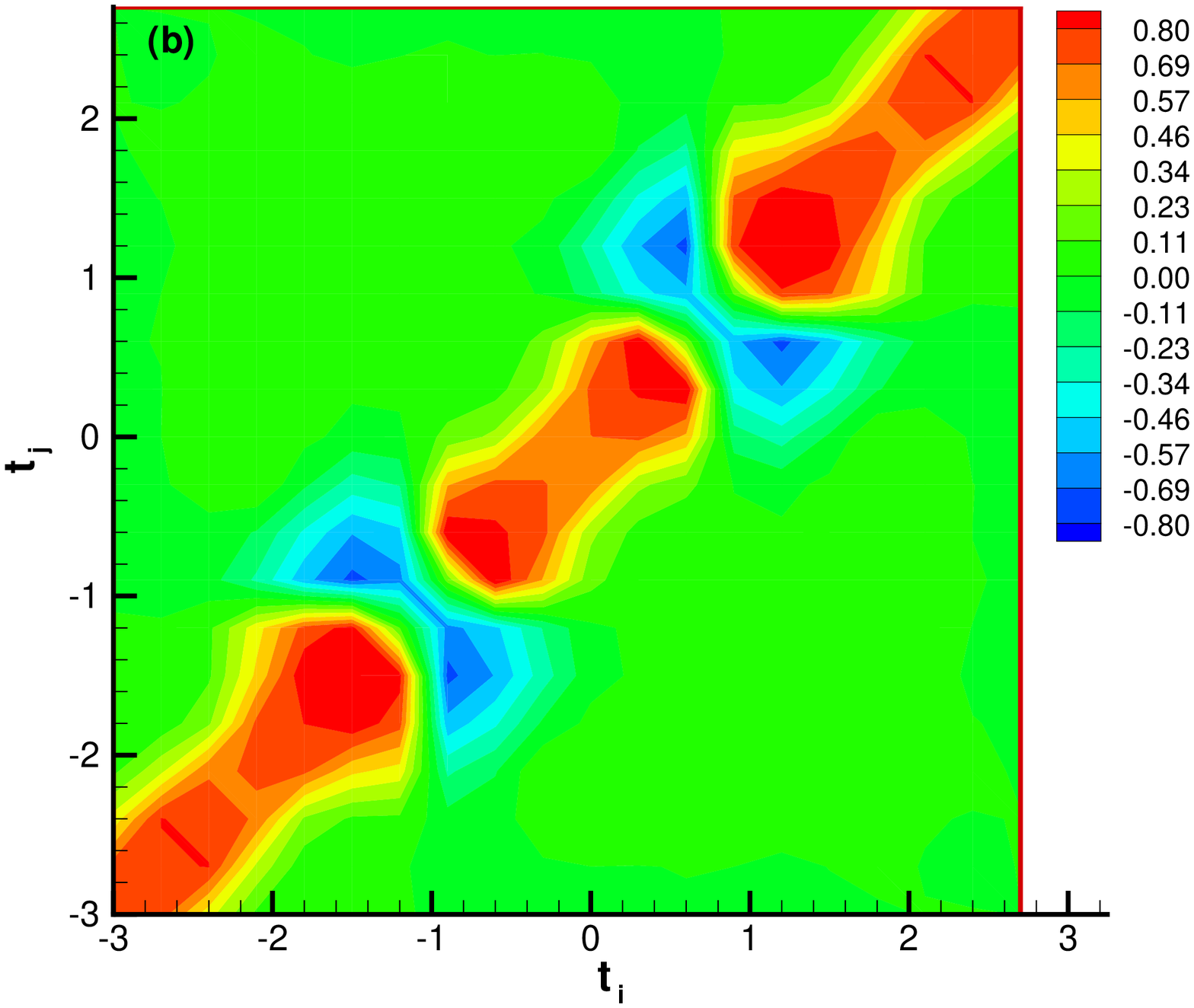} \end{center} 
\caption{A typical pattern of the time-domain photon-number 
correlations, $\protect\eta _{ij}$, for an out-of-phase two- soliton 
state in the cubic NLSE model (a), after $6$ normalized propagation 
distance; and for an in-phase two- soliton bound state in the CGLE 
model (b), after $0.5$ normalized propagation distance.}
\label{f-spectra} 
\end{figure}

In Fig. \ref{f-spectra}, we display the comparison of the 
\textit{time-domain} photon-number correlations for the two- soliton 
configuration in the conservative cubic NLSE model (a), and an in-
phase two-soliton bound state in the CGLE model (b). The correlation 
coefficients, $\eta _{ij}$, are defined through the normally-ordered 
covariance,
\begin{equation}
\eta _{ij}\equiv \frac{\langle :\Delta 
\hat{n}_{i}\Delta \hat{n}_{j}:\rangle}{\sqrt{\Delta 
\hat{n}_{i}^{2}\Delta \hat{n}_{j}^{2}}}~,
\label{C}
\end{equation} 
where $\Delta \hat{n}_{j}$ is the photon- number fluctuation in the 
$j$-th slot $\Delta t_{j}$ in the time domain, \[ \Delta 
\hat{n}_{j}=\int_{\Delta 
t_{j}}d\,t[U_{0}(z,t)\hat{u}^{\dag}(z,t)+U_{0}^{\ast 
}(z,t)\hat{u}(z,t)]. \] Here $\hat{u}(z,t)$ and $U_{0}(z,t)$ are the 
quantum-field perturbation and classical unperturbed solution, as 
defined above, and the integral is taken over the given time slot. 
In the NLSE model, there are two isolated patterns for the two-
solitons configuration, corresponding to the \textit{intra-pulse 
correlations} of individual solitons, see Fig. \ref{f-spectra}(a). 
However, it is obvious from Fig. \ref{f-spectra}(b) that there is a 
band of strong correlations between the two bound solitons in the 
quantum CGLE model. This strong-correlation band can be explained by 
the interplay between the nonlinearity, GVD, gain, and loss in the 
model. The balance between these features not only supports the 
classical stable bound state of the soliton, but also causes strong 
correlations between their quantum fluctuations.

In addition to the time-domain photon-number correlation pattern, we 
have also calculated a photon-number \textit{correlation parameter} 
between the two bound solitons, which is defined as 
\[ C_{12}=\frac{\langle :\Delta 
\hat{N}_{1}\Delta \hat{N}_{2}:\rangle}{\sqrt{\langle \Delta 
\hat{N}_{1}^{2}\rangle \langle \Delta \hat{N}_{2}^{2}\rangle }}. \] 
Here, $\Delta \hat{N}_{1,2}$ are perturbations of the photon-number 
operators of the two solitons, which are numbered (1,2) according to 
their position in the time domain.

\begin{figure}[tbp] \begin{center} 
\includegraphics[width=3.0in]{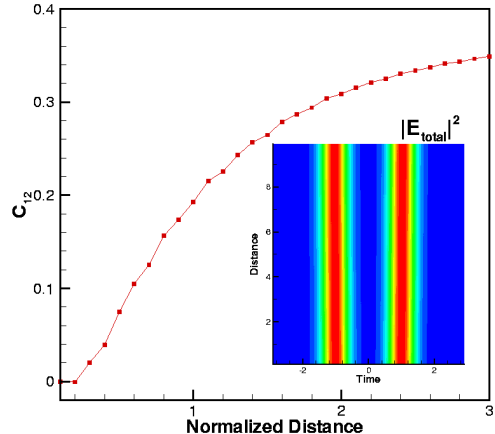} \end{center} 
\caption{The evolution of the photon-number correlation parameter, $C_{12}$, for two bound in-phase. The computations were carried out 
for the following values of parameters in the CGLE: $D=1$, 
$\protect\delta =-0.01$, $\protect\epsilon =1.8$, 
$\protect\beta=0.5$, $\protect\mu =-0.05$, and $\protect\nu =0$. The 
inset illustrates the stability of the underlying classical solution 
for the two- soliton bound state, by means of contour plots.}
\label{f-c12}
\end{figure}

Figure \ref{f-c12} shows the evolution of the photon-number 
correlation parameter in the two-soliton bound state. Initially, the 
classical laser statistics (coherent state) is assumed for each 
soliton, without correlation between them, $C_{12}\approx 0$. In the 
course of the evolution, the photon-number correlation between the 
two bound solitons gradually increases to positive values of 
$C_{12}$, and eventually it saturates about $C_{12}=0.36$. The 
inter- soliton correlation is induced and supported by the 
interaction between the solitons. In a conservative system, such as 
the NLSE model, nearly perfect photon-number correlations can be 
established if the interaction distance is long enough \cite{RK-en04}. In a non-conservative system, such as in the CGLE model, the action of the filtering, linear and nonlinear gain, and losses lead to the saturation of the photon-number correlation parameter. 
Thus, while the large quantum fluctuations in the output bound-
soliton pair will eclipse any squeezing effect, the correlated 
fluctuations between the bound soliton are predicted to be 
observable. 

\begin{figure}[tbp] \begin{center} 
\includegraphics[width=3.0in]{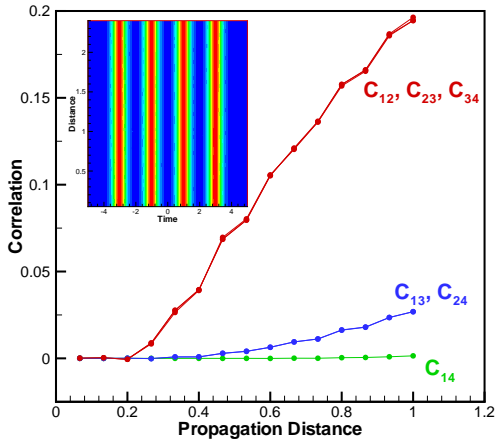} \end{center} 
\caption{The evolution of the photon-number correlation parameters, $C_{ij}$, for a bound complex (train) of four in-phase solitons in 
the CGLE model. The soliton's number ($1,2,3,4$) runs from left to 
right. Parameters are the same as in the case shown in Fig. 
(\protect\ref{f-c12}). The inset demonstrates the stability of the 
underlying classical solution for the four-soliton train.}
\label{f-c1234}
\end{figure}

The approach elaborated here for the study of the quantum-noise 
correlations between two bound solitons can be easily extended to 
multi-soliton bound states (soliton trains). Figure \ref{f-c1234} 
shows the photon-number correlation parameters, $C_{ij}$, in a train 
of four equal-separated in-phase bound solitons. Again, the photon-number fluctuations are initially uncorrelated between the solitons. 
As could be expected, we find that soliton pairs with equal 
separations have practically identical correlation coefficients, 
i.e. , $C_{12}(z)\approx C_{23}(z)\approx C_{34}(z)$, and 
$C_{13}(z)\approx C_{24}(z)$. Obviously, when the interaction 
between the soliton trains is stronger, as the separation between 
them is smaller, the values of the correlation parameter are larger. 
Note that the correlation coefficient for the most separated pair in 
the train, $C_{14}$, grows very slowly with $z$. Eventually, all of 
the curves of $C_{ij}(z)$ saturate due to the dissipative effects in 
the CGLE model. 

In conclusion, in this paper we have extended the concept of quantum 
fluctuation around classical optical solitons to the non-conservative model based on the CGLE (complex Ginzburg-Landau 
equation) with the cubic-quintic nonlinearity. Applying the known 
back-propagation method to the linearized equations that govern the 
evolution of the fluctuations, we have numerically calculated the 
photon-number correlations and the effective correlation coefficient 
for pairs of bound solitons, as well as for multi-soliton bound 
complexes (trains). We have demonstrated that, unlike the two-soliton configuration in the conservative NLSE model, there is a 
band of strong quantum correlation in the bound-soliton pair. While 
the dissipative effects in the CGLE model will totally suppress the 
generation of squeezed states from bound solitons, as one might 
expect, there still exists a certain degree of correlations between 
photon-number fluctuations around the stable bound-soliton pairs and 
trains. Recently, experimental progress in the study of various 
quantum properties of solitons in optical fibers has been reported 
\cite{Silberhorn01, Silberhorn02, Glockl03, Konig02}, which opens 
the way to observe effects predicted here. Besides that, the photon-
number-correlated soliton pairs and trains, predicted in this work, 
may offer new possibilities to generate multipartite entangled 
sources for applications to quantum communications and computation.

\end{document}